\documentclass[]{spie}  

\usepackage{amsmath,amsfonts,amssymb}
\usepackage{graphicx}
\usepackage[colorlinks=true, allcolors=blue]{hyperref}

\title{PYRA and MYTHRA: new statistical analysis for image reconstruction}

\author[a]{Julien Drevon}
\author[b]{Margaux Abello}
\author[b]{Florentin Millour}
\author[b]{Anthony Meilland}
\author[b]{Armando Domiciano de Souza}

\affil[a]{European Southern Observatory, Alonso de Córdova 3107, Vitacura, Santiago, Chile}
\affil[b]{Université Côte d'Azur, Observatoire de la Côte d'Azur, CNRS, Lagrange, France}

\authorinfo{Further author information: \\J.D.: E-mail: julien.drevon@eso.org}

\begin{document} 
\maketitle

\begin{abstract}
Image reconstruction in optical interferometry remains a challenging inverse problem due to sparse Fourier sampling and the strong dependence of the reconstructed images on user-defined parameters. We present \texttt{PYRA} (Python for MiRA) and \texttt{MYTHRA} (Mean Astrophysical Images with PYRA), two tools designed to improve the robustness and reproducibility of interferometric imaging. \texttt{PYRA} automatically generates large ensembles of \texttt{MiRA} reconstructions by exploring a wide range of reconstruction parameters, while \texttt{MYTHRA} statistically analyzes the resulting images to identify the most consistent solutions and derive a final averaged image with associated uncertainty estimates. The methodology is validated using the blind datasets of the 2024 Interferometric Imaging Contest, where our reconstruction strategy achieved the best overall score. We further apply the method to VLTI/MATISSE observations of the triple AGB system $\pi^1$~Gru and the A[e] supergiant 3~Pup. These results demonstrate that statistical analysis of large reconstruction ensembles provides a powerful framework for assessing image reliability and reducing reconstruction artifacts in optical interferometry
\end{abstract}

\keywords{optical interferometry, image reconstruction, statistical imaging, inverse problems, interferometric data analysis, high angular resolution astronomy}

\section{Introduction}\label{sec:intro} 

Optical interferometry enables astronomical observations to achieve angular resolutions far beyond those attainable with current single-aperture or segmented-mirror telescopes. Even the largest segmented telescope currently under construction, the Extremely Large Telescope (39~m aperture), cannot reach the angular resolution provided by interferometric facilities such as the Very Large Telescope Interferometer (VLTI) located at Cerro Paranal in Chile. This is because the angular resolution of a single-dish telescope is fundamentally limited by the diameter of its primary mirror, whereas in an interferometer it is determined by the separation between the telescopes, also known as the baseline. By combining the light collected by multiple telescopes, the VLTI can achieve an angular resolution equivalent to that of a telescope with an effective aperture of up to about 200~m. This allows observations to probe structures on milliarcsecond (mas) angular scales in the mid-infrared.

However, unlike single-dish telescopes, the measured quantities in optical interferometry do not correspond directly to the intensity distribution of the observed object. Because the light collected by separate telescopes is combined to produce interference fringes, the observables are sampled in the Fourier domain rather than in the image plane. In practice, interferometers measure quantities such as the visibility, which corresponds to the contrast of the interference fringes and provides information about geometrical structures in the brightness distribution, and the closure phase, defined as the sum of the measured phases along a triangle of baselines. The closure phase is particularly important because it cancels the phase perturbations introduced by the atmosphere, allowing the intrinsic phase information of the observed object to be partially recovered.

Recovering the image of the observed object from interferometric observables therefore constitutes an inverse problem. Two main approaches are commonly used: analytical modeling based on simple parametric representations of the object, or image reconstruction techniques. Analytical modeling relies on fitting a predefined geometrical model to the data, typically through the minimization of a $\chi^2$ function. Image reconstruction, on the other hand, searches for the image that best reproduces the observables without assuming a specific parametric model.

Several image reconstruction software packages are currently used by the optical interferometry community, most of which adopt a Bayesian framework, including \texttt{MiRA} \cite{Thiebaut2008} , \texttt{SQUEEZE} \cite{Baron2010} , \texttt{BSMEM} \cite{Baron2008} , \texttt{IRBis} \cite{Hofmann2014} , and \texttt{SPARCO} \cite{Kluska2014} . One of the key challenges in image reconstruction is the determination of an appropriate value for the hyperparameter $\mu$, which defines the balance between fitting the data and enforcing the regularization. To date, no widely adopted automatic method has been established in the literature to reliably determine this parameter and ensure robust reconstructions.

A common strategy used by the community to validate reconstructed structures is to compare the results obtained with several independent reconstruction algorithms and assess whether they converge toward similar images. In this work, we propose a different approach. Instead of relying on multiple reconstruction algorithms, we exploit a single reconstruction software, namely \texttt{MiRA} (Multi-aperture Image Reconstruction Algorithm \cite{Thiebaut2008}), to explore a large region of the reconstruction parameter space. Thousands of reconstructions are generated while varying parameters such as the regularization weight $\mu$, the field of view (FoV), and the pixel size $\theta$. A statistical analysis of this ensemble of reconstructed images is then performed in order to identify the most consistent subset of solutions and derive a robust final image. This methodology is implemented through the tools \texttt{PYRA} and \texttt{MYTHRA}.

\section{Interferometric imaging framework}\label{sec:imaging}  

\subsection{Fourier sampling in optical interferometry}

Each pair of telescopes observes the sky with a specific projected baseline length and position angle, which corresponds to sampling a single spatial frequency of the object in the Fourier domain. As a result, every telescope pair probes one point in what is commonly referred to as the $(u,v)$-plane. In contrast, a single-dish telescope measures spatial frequencies continuously from zero up to a maximum value set by the diameter of its aperture.

Optical interferometers therefore provide only sparse sampling of the Fourier plane. To improve the reconstruction of the observed object, it is necessary to vary both the baseline length and its orientation between the telescopes, allowing different spatial frequencies to be measured. By combining observations obtained with multiple baselines and position angles, it becomes possible to increase the coverage of the $(u,v)$-plane and recover more complete information about the structure of the target.

\subsection{Image reconstruction as an inverse problem}

Image reconstruction aims to recover the brightness distribution of the observed object that is consistent with the interferometric observables. This is generally achieved by minimizing a global cost function of the form
\[
f_{\mathrm{tot}} = \chi^2 + \mu \times f_{\mathrm{reg}} \;,
\]
where $\chi^2$ represents the data fidelity term, $f_{\mathrm{reg}}$ is the regularization function encoding prior assumptions about the object (such as smoothness or compactness), and $\mu$ is the hyperparameter controlling the relative weight between the data term and the regularization term.

\begin{figure}
    \centering
    \includegraphics[width=1\linewidth]{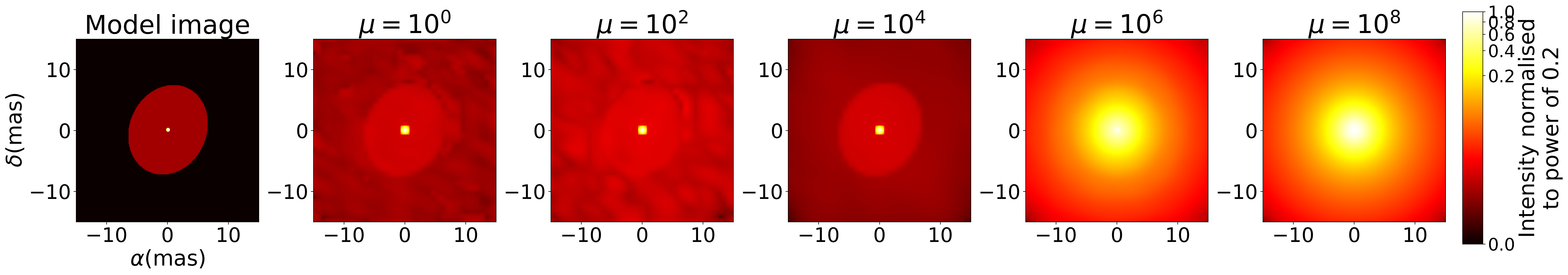}
    \caption{Five images reconstructed with the software \texttt{MiRA} using identical settings, besides the hyperparameter $\mu$ that varies logarithmically from $10^0$ to $10^8$. The reconstructions have been obtained with the compactness regularization function, parametrized by $\gamma = 5$~mas, over a FoV of 30~mas, and a pixel size of 0.1~mas/pix.}
    \label{fig:mu_influence}
\end{figure}

The choice of the hyperparameter $\mu$ plays a critical role in the reconstruction process, as it determines the balance between fitting the data and enforcing the prior information  (Fig.~\ref{fig:mu_influence}). An excessively small value may lead to noisy reconstructions dominated by the data, while an excessively large value may produce overly smooth images dominated by the regularization. Determining an optimal value for this parameter therefore remains a key challenge in interferometric image reconstruction.

\section{PYRA: large-scale image reconstruction generation}\label{sec:PYRA} 

\texttt{PYRA} (Python for MiRA; \url{https://github.com/jdrevon/PYRA/tree/main}) is a Python wrapper developed to facilitate and optimize the use of the \texttt{MiRA} software for interferometric image reconstruction. The main objective of \texttt{PYRA} is to automate the exploration of the reconstruction parameter space in order to generate a large ensemble of reconstructed images compatible with the interferometric observables.

The tool randomly samples several key parameters, including: i) the pixel size, ii) the FoV, iii) the hyperparameter that controls the weight of the regularization term, and iv) regularization-specific parameters, such as $\gamma$ for the compactness regularization or $\tau$ for the hyperbolic regularization. The sampling intervals (one per key parameter mentioned) are defined using physically motivated limits derived from the interferometric resolution and the FoV of the observations.

The pixel size $\theta$ is randomly selected within an interval defined by the interferometric resolution, typically between $\lambda_{\min}/6 B_{\max}$ and $\lambda_{\min}/2 B_{\max}$, ensuring both adequate sampling and avoiding aliasing effects. Similarly, the FoV is chosen within a range derived from the interferometric FoV, typically between $1\, \times \mathrm{FoV}$ and $2 \times \,\mathrm{FoV}$. The hyperparameter $\mu$ is explored over several orders of magnitude, generally between $10^3$ and $10^9$, in order to probe a wide range of balances between the data fidelity and regularization terms. The number of iterations is set by default to approximately 50, which appears to be a good compromise between underfitting the data and entering an overfitting regime. It also increases the generation speed of the images. 

By combining random sampling of these parameters with automated batch execution of \texttt{MiRA}, \texttt{PYRA} allows the generation of thousands of reconstructed images. This ensemble of reconstructions provides a comprehensive exploration of the solution space, which can then be statistically analyzed to identify robust structures in the reconstructed object.

\section{MYTHRA: statistical analysis of reconstructed images}\label{sec:MYTHRA} 

\texttt{MYTHRA} (Mean Astrophysical Images with PYRA; \url{https://github.com/jdrevon/MYTHRA/tree/main}) is a Python tool developed to analyze the reconstructed images generated by \texttt{PYRA}. Its purpose is to perform a statistical analysis of the ensemble of reconstructions, select the most relevant subset of images, and produce a final robust image through averaging.

To achieve this, the code first identifies the most relevant subset of reconstructions using the so-called L-curve criterion, which characterizes the balance between the data fidelity and regularization terms. The L-curve provides a first indication of whether the reconstruction reaches a meaningful compromise, converging toward a consistent and plausible physical solution. When the hyperparameter $\mu$ is small, the minimization of the total function $f_{\mathrm{tot}}$ is dominated by the data term, leading to small $\chi^2$ values. As $\mu$ increases, the regularization term progressively dominates the minimization process, and the $\chi^2$ term begins to increase significantly. This regime corresponds to overly regularized solutions that are no longer consistent with the data and should therefore be avoided. The divergence point of the L-curve is automatically estimated by binning the reconstruction results and identifying the transition between data-dominated and regularization-dominated regimes.

Once the divergence point is determined, a subset of reconstructions located within a user-defined window before this region is selected. In some cases, additional filtering is required due to the presence of outliers. Since the initial selection is based on the $\mu$ values, it does not necessarily prevent outliers (i.e reconstructions with high $\chi^2$ values) from being included in the subset. To address this issue, an optional $\chi^2$ filtering can be applied by rejecting images whose $\chi^2$ values fall outside user-defined boundaries.

After this filtering step, the selected set of reconstructed images must be homogenized before computing the averaged image, which is obtained as the mean of the set. Indeed, the reconstructions may have been obtained with different pixel sizes or FoVs, resulting in images with different effective resolutions. To ensure consistency, all selected images are resampled onto a common pixel grid. This is achieved by re-running \texttt{MiRA} using each reconstructed image as the initial guess while fixing the FoV and $\theta$ to the smallest FoV and to half the smallest pixel size from the sample, respectively. This procedure ensures that all images share the same spatial resolution while preserving consistency with the interferometric observables and maintaining a robust global $\chi^2$ value.

The last step before computing the averaged image from the selected subset is to check for the presence of potential undetected outliers. In some cases, an image may present a very good $\chi^2$ value while still being dominated by reconstruction artifacts. This can occur when the algorithm places flux in poorly sampled regions of the $(u,v)$-plane coverage, producing spurious structures that do not correspond to the true object morphology.

To mitigate this effect, we implemented an iterative averaging procedure. Images are added sequentially to the subset, and at each step the averaged image is recomputed. The impact of including each additional image is evaluated by monitoring the variation of the $\chi^2$ computed on the squared visibilities ($\chi^2_\mathrm{V^2}$) and on the closure phases ($\chi^2_\mathrm{CP}$) independently. If the inclusion of a given image produces a significant degradation of either metric, the image is considered a potential outlier and can be excluded from the final subset. The acceptance threshold is defined by the user based on the statistical performance of the $\chi^2$ values within the subsample.

Once the contribution of each image to the global $\chi^2$ is well constrained, the final step consists of computing the averaged image from the selected subset of reconstructions. In addition, pixel-wise error maps are derived to characterize the statistical dispersion of the solutions. This statistical approach allows robust structures that consistently appear across the reconstructions to be identified, while reducing the influence of artifacts associated with specific reconstruction parameters.

\section{Validation on simulated data}\label{sec:sim_data} 

To validate the statistical imaging framework implemented in \texttt{PYRA} and \texttt{MYTHRA}, we applied the method to the simulated datasets distributed for the 2024 Interferometric Imaging Contest (I2C@2024 \cite{Millour2024}). The contest provided blind chromatic interferometric datasets generated from realistic VLTI simulations, including instrumental noise and calibration biases, for two objects: a Wolf-Rayet star with spirals and a multi-gaped circumstellar disk containing embedded companions.

Our contribution consisted in reconstructing wavelength-dependent image cubes with \texttt{MiRA} through a large exploration of the reconstruction parameter space, and in combining the resulting images statistically using the \texttt{PYRA}/\texttt{MYTHRA} workflow. For each dataset, reconstructions were produced over a wide range of hyperparameters, pixel scales, and FoVs, and the final images were obtained from the statistically selected subset of acceptable solutions. This strategy differs from a traditional single-reconstruction approach by explicitly quantifying the stability of reconstructed structures against reconstruction parameters.

The contest organizers evaluated all submissions using a translation- and flux-invariant L1 image metric, after matching the effective spatial resolution and spectral sampling of the reconstructed images to the reference models. The score was computed independently for each spectral channel and then averaged over wavelength.

Using this evaluation framework, our \texttt{MiRA}-based submission achieved the best overall score of the contest, and was therefore selected as the winning contribution of I2C@2024. The reconstructions successfully recovered the main morphological features of both the spiral object and the disk, while the statistical combination of reconstructions reduced wavelength-dependent artifacts and improved the robustness of the recovered structures.

This blind validation on realistic simulated interferometric datasets demonstrates that the \texttt{PYRA}/\texttt{MYTHRA} approach can recover reliable image structures while simultaneously providing a natural framework for estimating reconstruction uncertainties from the dispersion of acceptable solutions.

\section{Application to real observations}\label{sec:real_data} 
In this section we present two applications of \texttt{PYRA} and \texttt{MYTHRA} to real interferometric observations. These examples use different regularization methods and illustrate two evolved stellar systems representative of different evolutionary pathways: an asymptotic giant branch (AGB) star and a supergiant star.

\subsection{The triple system $\pi^1$~Gru with an S-type AGB star}
Within the framework of the MATISSE BIN-AGB Large Program (ESO ID 108.22E9), the triple system $\pi^1$~Gru was observed with VLTI/MATISSE in both $L$ and $N$ bands. The primary component, $\pi^1$~Gru~A, is an evolved S-type AGB star, while $\pi^1$~Gru~B is a distant companion located at a separation of about 2~arcsec. A third companion, $\pi^1$~Gru~C, was more recently detected with ALMA at a separation of about 40~mas from the primary within the framework of the ATOMIUM Large Program.

The high angular resolution provided by VLTI/MATISSE makes these observations particularly suitable for investigating the close environment of the system. In particular, the dataset allow us to probe the possible interactions between $\pi^1$~Gru~A and the close companion $\pi^1$~Gru~C. These observations also provide insight into how molecular and dust formation processes in the circumstellar environment are affected by binarity. Studying these small-scale structures is essential to connect them with the larger-scale morphology revealed by instruments such as VLT/SPHERE and ALMA.

Due to the complete $(u,v)$-plane obtained at the end of the MATISSE Large Program, the interferometric observations were then processed to reconstruct images using \texttt{PYRA} and \texttt{MYTHRA}. For both $L$ and $N$ band data, we explored a range of reconstruction hyperparameters in \texttt{PYRA}. The settings used for the image reconstruction are summarized below:
\begin{itemize}
    \item Number of individual reconstructions: 1\,000 images;
    \item $\theta$ range: $\left[\frac{1}{3} \times \frac{\lambda_{\min}}{2 B_{\max}},\; 1 \times\frac{\lambda_{\min}}{2 B_{\max}} \right]$;
    \item FoV range: $\left[1.0 \times \frac{\lambda_{\max}}{B_{\min}},\; 2.0 \times \frac{\lambda_{\max}}{B_{\min}} \right]$;
    \item $\mu$ search range: $\left[10^{2},\; 10^{9}\right]$.
\end{itemize}
We applied the compactness regularization, which favors solutions where the flux is concentrated within a limited area defined by $\gamma$. This angular scale factor can be interpreted as the full width at half maximum (FWHM) of a Gaussian profile and was constrained as follows:
\begin{itemize}
    \item $\gamma$ range: $\left[0.5 \times \frac{\lambda_{\max}}{B_{\min}},\; 1.0 \times \frac{\lambda_{\max}}{B_{\min}} \right]\;.$
\end{itemize}
In practice, for this astrophysical system, we obtained better reconstructions using the compactness regularization compared to the hyperbolic regularization.

After running \texttt{PYRA}, we obtained the L-curve for both reconstructed images in the $L$ and $N$ bands, as shown in Fig.~\ref{fig:l-curve-piGru_LN}. In both cases, the L-curves obtained are smooth and well defined, with negligible scatter and $\chi^2$ values close to unity. From these L-curves, \texttt{MYTHRA} automatically identifies the divergence point, shown as a red dashed line, and selects the subset of images used to initiate the statistical analysis, indicated by the colored points.

Fig.~\ref{fig:images_LN_piGru} presents the final image products obtained with \texttt{MYTHRA}. The top panel shows the reconstructed image in the $L$ band together with the comparison between the model and the observables. The same type of analysis is shown for the $N$ band in the bottom panel.

Both the $L$ and $N$ band images reveal structures of interest that are discussed in detail in a dedicated publication \cite{Drevon2026} . In summary, the $L$ band image shows a deformation of the stellar atmosphere in the direction of the expected location of the binary companion at the time of the VLTI/MATISSE observations. The morphology suggests a mass transfer between the central AGB star and the companion. In addition, a disk like structure appears around the expected position of the companion at a separation of about 40~mas from the AGB star. The $N$ band image shows a clear circumcompanion environment with a tail that appears to be the origin of the larger scale arms observed farther from the system.

The robustness of the structures reconstructed with \texttt{PYRA} and \texttt{MYTHRA} is also supported by the superposition of the VLTI/MATISSE images with VLT/SPHERE observations, which show a strong spatial agreement between the detected structures (Fig.~\ref{fig:composite_piGru}). The VLTI/MATISSE reconstructions therefore bridge the gap between the small scale structures and the larger scale morphology of the system, providing a clearer view of the influence of the companion on the dust and molecular distribution in the close environment of the star. 

\subsection{The binary system 3~Pup with an A-type B[e] supergiant}
Within the framework of the MATISSE B[e] stars survey, 3~Pup was observed with VLTI/MATISSE as part of the Guaranteed Time Observations program (ESO IDs 0104.D-0669, 0104.D-0015, 0104.D-0554, and 112.25C5), simultaneously in the $L$, $M$, and $N$ bands. This binary system, also known as $l$~Pup or HD~62623, stands out as the brightest known Galactic object exhibiting the B[e] phenomenon. The primary component is an A-type supergiant embedded in a gaseous and dusty circumbinary disk, while the secondary is a low-mass spectroscopic companion located at a close angular separation of about 1.76~mas. 

Given the presence of a dense circumstellar environment and the potentially complex morphology resulting from the dynamical influence of the binary companion, the high angular resolution provided by VLTI/MATISSE makes these observations particularly well suited for investigating the structured disk of 3~Pup close to the sublimation radius, at angular scales of 1.5~mas at 3.0~$\mu$m and 6.2~mas at 12~$\mu$m. 

Because of the well-distributed $(u,v)$-plane, the interferometric observations of 3~Pup were processed to reconstruct images using \texttt{PYRA} and \texttt{MYTHRA}. For each spectral band offered by VLTI/MATISSE, identified by its central wavelength $\lambda_0$, we generated a set of reconstructions with \texttt{PYRA} using the following settings:
\begin{itemize}
    \item Number of reconstructions: 3\,360 images;
    \item $\theta$ range: $\left[0.5 \times \frac{\lambda_{0}}{2 B_{\max}},\; 1\times \frac{\lambda_{0}}{2 B_{\max}}\right]$;
    \item FoV range: $\left[0.5 \times \frac{\lambda_{0}}{B_{\min}},\; 1.5 \times \frac{\lambda_{0}}{B_{\min}}\right]$;
    \item $\mu$ search range: $\left[10^{0},\; 10^{8}\right]$.
\end{itemize}
We applied the hyperbolic regularization, an edge-preserving smoothness prior that favors solutions with sharp features and is efficient in reconstructing extended structures, such as a dusty inner rim. This regularization method relies on a threshold parameter, $\tau = \left(\theta / \text{FoV}\right)^2$, which estimates the typical absolute difference between neighboring pixels. Since $\tau$ is defined according to the spatial characteristics of each reconstructed image, its value is systematically adapted. The parameter was sampled in \texttt{PYRA} as follows:
\begin{itemize}
    \item $\tau$ range: $\left[ 10^{-4} \times \tau ,\; 10^{+1} \times \tau \right]$.
\end{itemize} 
For this object, we obtained similar averaged images using compactness regularization. 

After running \texttt{PYRA}, we obtained the L-curves for the reconstructed images in the $L$, $M$, and $N$ bands, as shown in Figs.~\ref{fig:l-curve-lPup_L} and \ref{fig:l-curve-lPup_MN}. \texttt{MYTHRA} then automatically identified the divergence point following the same methodology as for the analysis of $\pi^1$~Gru. Figs.~\ref{fig:images_LM_lPup} and \ref{fig:images_N_lPup} present the final averaged images, revealing unexpected large-scale structures beyond the dusty inner rim, which are supported by the derived pixel-wise error maps displayed in Figs.~\ref{fig:confmaps_lPup_Lband} and \ref{fig:confmaps_lPup_MNbands}. A complementary study combining these \texttt{MYTHRA} images with multi-component geometric modeling, independent reconstructions with \texttt{SPARCO}, and a hydrodynamical simulation was published to propose a physical interpretation for the discovered asymmetries \cite{Abello2025} . In summary, the morphology and radial extent of these large-scale structures located to the south-east and the north-west regions of 3~Pup's dusty disk are best interpreted as tidally induced spiral density waves driven by the central binary. 

Moreover, the elongated structures identified individually in each spectral band are also apparent in the composite image displayed in Fig.~\ref{fig:composite_lPup}, revealing a clear stratification of the circumstellar emission with wavelength. The innermost region, dominated by the bright yellow-white central source, reflects the superposition of all three bands and corresponds to the unresolved or marginally resolved stellar binary and its immediate hot gaseous environment. Moving outward, the emission progressively transitions from blue-dominated, tracing the hotter and more compact $L$ band emission close to the dust sublimation zone, to green in the intermediate region corresponding to the $M$ band, and finally to red at the largest angular scales, where the cooler $N$-band emission from the outer dusty disk dominates. This chromatic gradient is consistent with the expected temperature stratification of a Keplerian circumbinary disk, in which the dust temperature decreases with distance from the central heating source. The VLTI/MATISSE reconstructions therefore provide a clearer view of the influence of the companion on the dust distribution in the close environment of the supergiant star.

\section{Conclusion}\label{sec:conclu} 

In this work, we introduced \texttt{PYRA} and \texttt{MYTHRA}, two complementary tools developed to improve the robustness and reproducibility of interferometric image reconstruction. Instead of relying on comparisons between different reconstruction algorithms, our approach explores a large region of the reconstruction parameter space within a single reconstruction framework, \texttt{MiRA}, combined with a statistical analysis of thousands of reconstructed images.

\texttt{PYRA} enables the automated generation of large ensembles of reconstructions by randomly sampling key reconstruction parameters, such as the hyperparameter, the FoV, the pixel size, and regularization-specific quantities. \texttt{MYTHRA} then performs the statistical selection and homogenization of the reconstructed images, identifies potential outliers, and derives a final averaged image together with associated uncertainty maps. This methodology allows robust astrophysical structures that consistently appear across the reconstructions to be isolated, while reducing the influence of reconstruction artifacts linked to specific parameter choices.

The application of the method to VLTI/MATISSE observations of $\pi^1$~Gru and 3~Pup demonstrates the capability of the approach to recover physically meaningful structures in complex interferometric datasets. The averaged images reveal asymmetric morphologies and circumstellar or circumcompanion features that are likely shaped by companion-driven interactions, and whose reliability is supported by independent observations, complementary modeling, or independent image-reconstruction algorithms. Beyond these specific astrophysical cases, the methodology presented here provides a general framework for statistical interferometric imaging. It offers a practical way to evaluate the stability of reconstructed features and to quantify reconstruction uncertainties in a more systematic manner. 

Future developments will focus on extending the approach to chromatic image reconstruction and to other image reconstruction software, improving the automatic identification of optimal reconstruction subsets by exploring the hypothesis that the L-curve depends on multiple reconstruction parameters, and integrating additional statistical diagnostics to further strengthen the reliability of interferometric imaging analyses.

\section{Acknowledgments}

J.D. and M.A. acknowledge the support from the ESO Fellowship Programme and the Early-Career Scientific Visitor Programme at ESO Chile, respectively. This work made use of the Jean-Marie Mariotti Center SearchCal service (\url{http://www.jmmc.fr/searchcal}), co-developed by LAGRANGE and IPAG.
Part of this work has received funding from the French Agence Nationale de la Recherche (ANR), under grant MASSIF (\url{https://anr.fr/Projet-ANR-21-CE31-0018}).
J.D acknowledge the support from the JMMC group and particularly Eric Thiébaut for their time to talk about the MiRA software and their support for the creation of PYRA and MYTHRA.

\newpage
\bibliography{spie1414818} 
\bibliographystyle{spie} 

\newpage
\appendix{}

\section{Statistical image reconstruction results on $\pi^1$~Gru}
\begin{figure}[htbp!]
    \centering
    \includegraphics[width=0.70\linewidth]{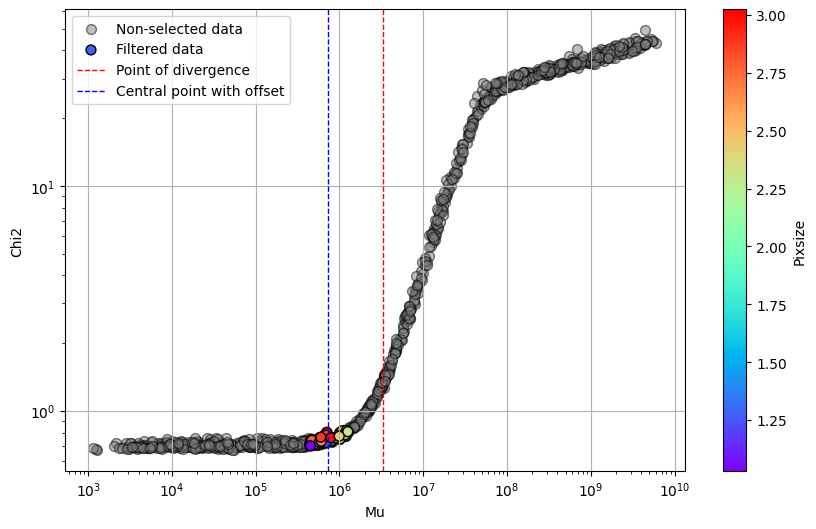}
    \includegraphics[width=0.70\linewidth]{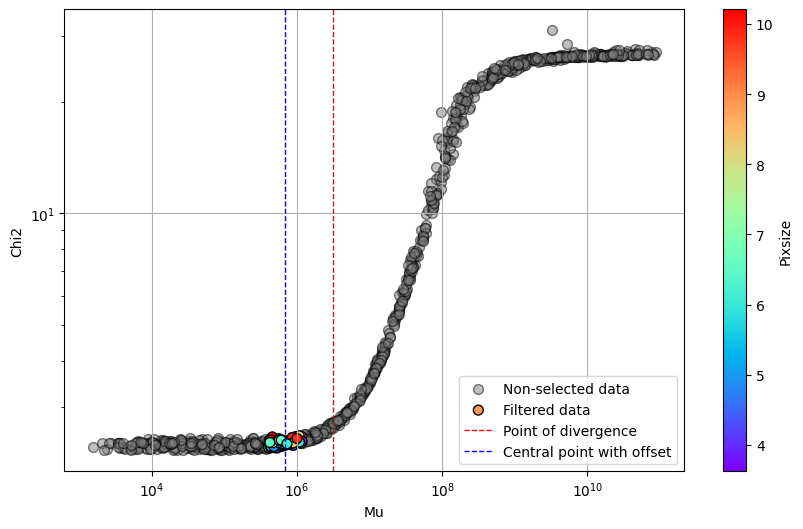}
    \caption{L-curves obtained for the reconstructed images of $\pi^1$~Gru using VLTI/MATISSE data. The colored points correspond to the solutions selected by \texttt{MYTHRA}. Top: in the $L$ band. Bottom: in the $N$ band.}
    \label{fig:l-curve-piGru_LN}
\end{figure}

\begin{figure}[htbp!]
    \centering
    \includegraphics[width=0.8\linewidth]{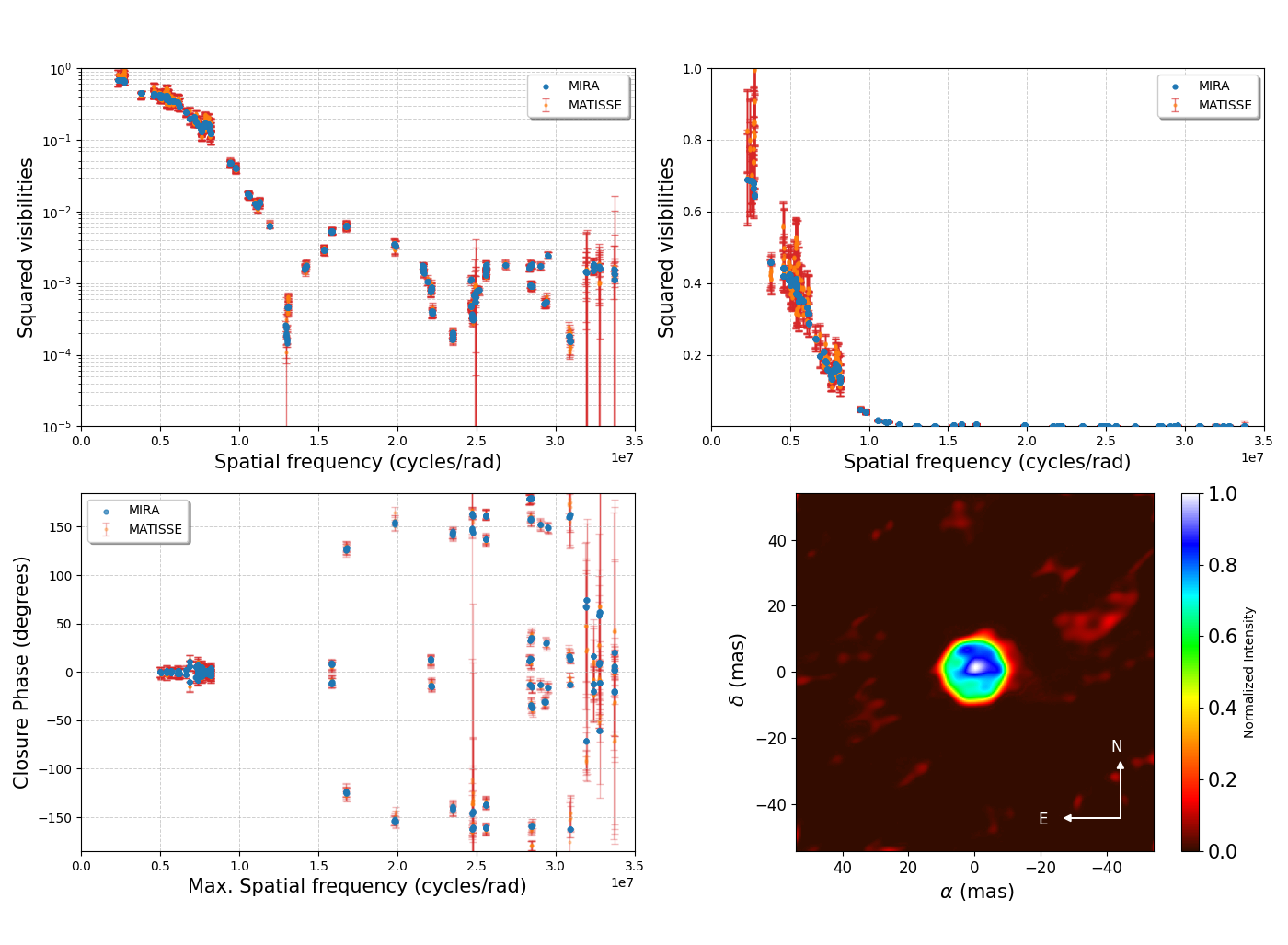}
    \includegraphics[width=0.8\linewidth]{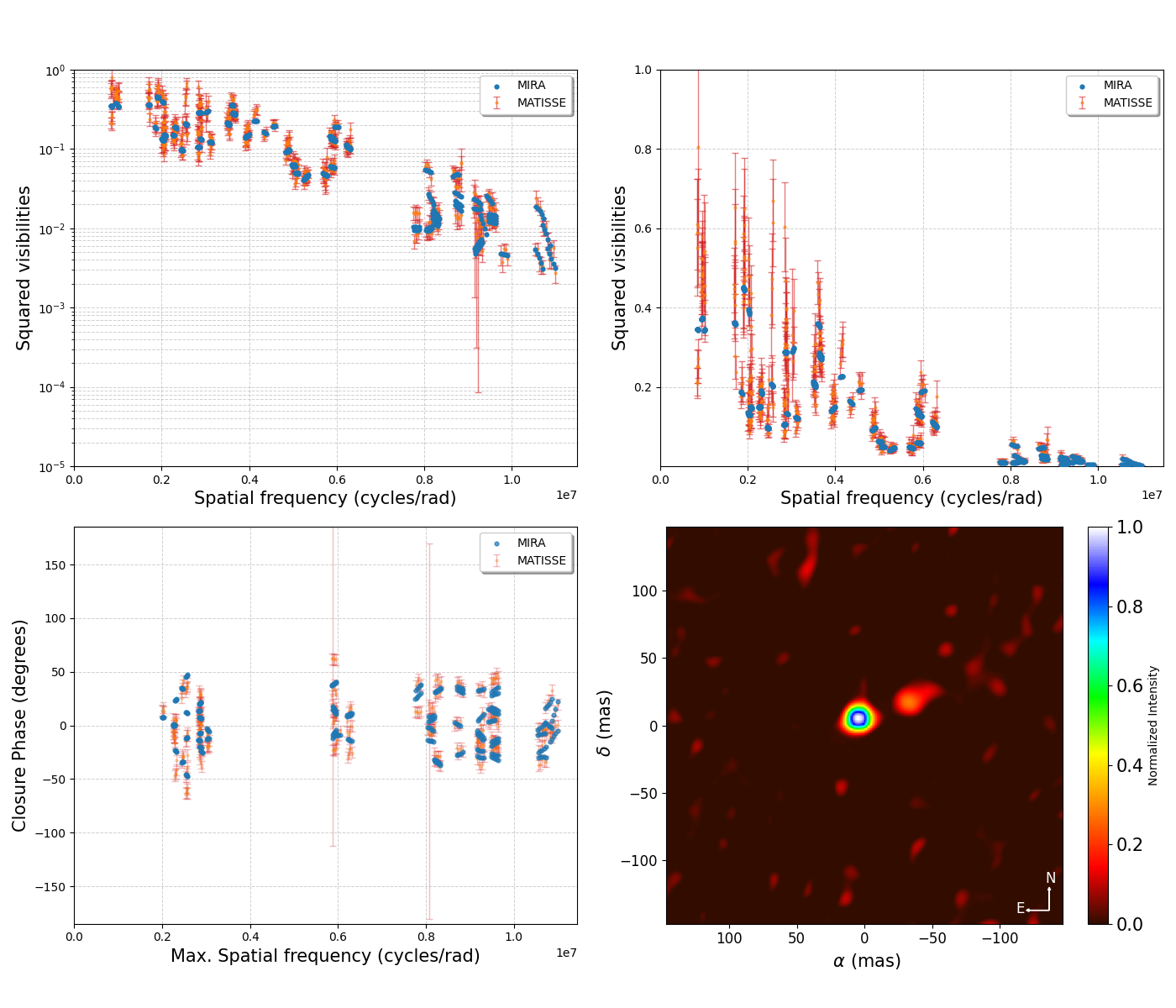}
    \caption{Images of $\pi^1$~Gru reconstructed with \texttt{MYTHRA}, including the comparison with the observed squared visibilities (in logarithmic and linear scales) and the observed closure phases. Top: in the $L$ band. Bottom: in the $N$ band.}
    \label{fig:images_LN_piGru}
\end{figure}

\begin{figure}[htbp!]
    \centering
    \includegraphics[width=0.7\linewidth]{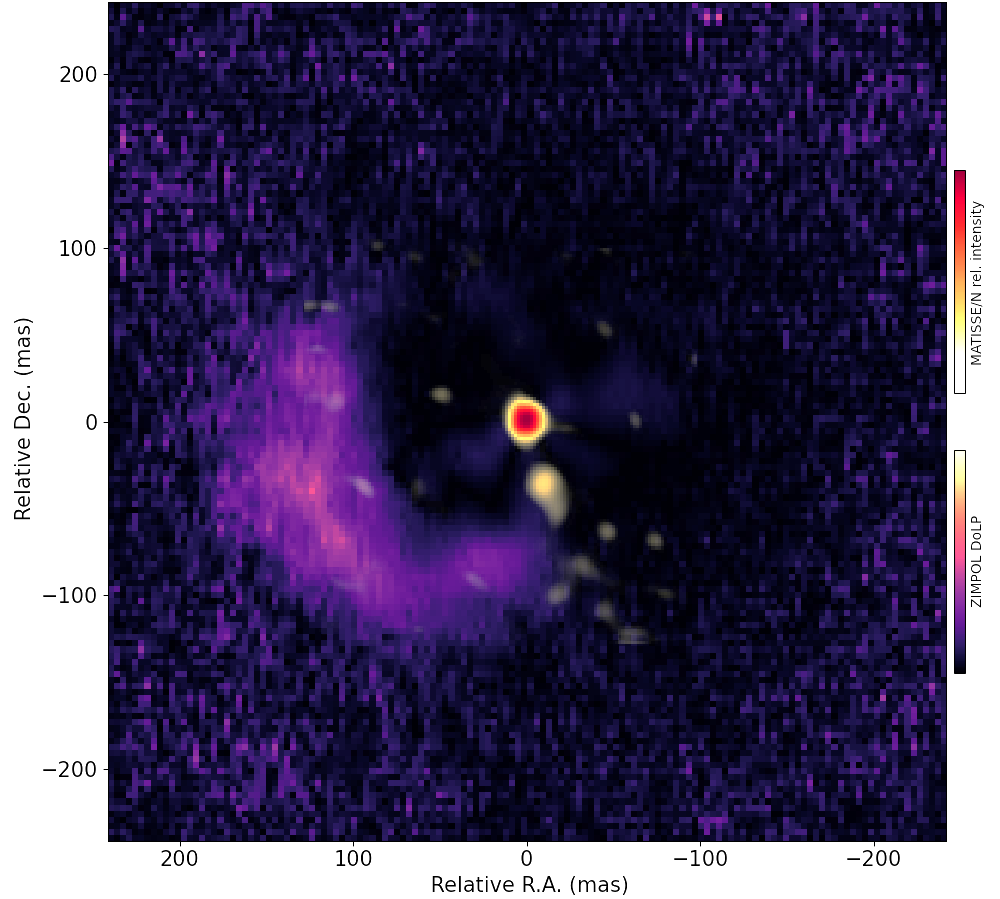}
    \caption{The composite image of $\pi^1$~Gru combines VLT/SPHERE visible data (in purple) with VLTI/MATISSE $N$ band data (in red/yellow).}
    \label{fig:composite_piGru}
\end{figure}

\newpage

\section{Statistical image reconstruction results on 3~Pup}

\begin{figure}[htbp!]
    \centering
    \includegraphics[width=0.75\linewidth]{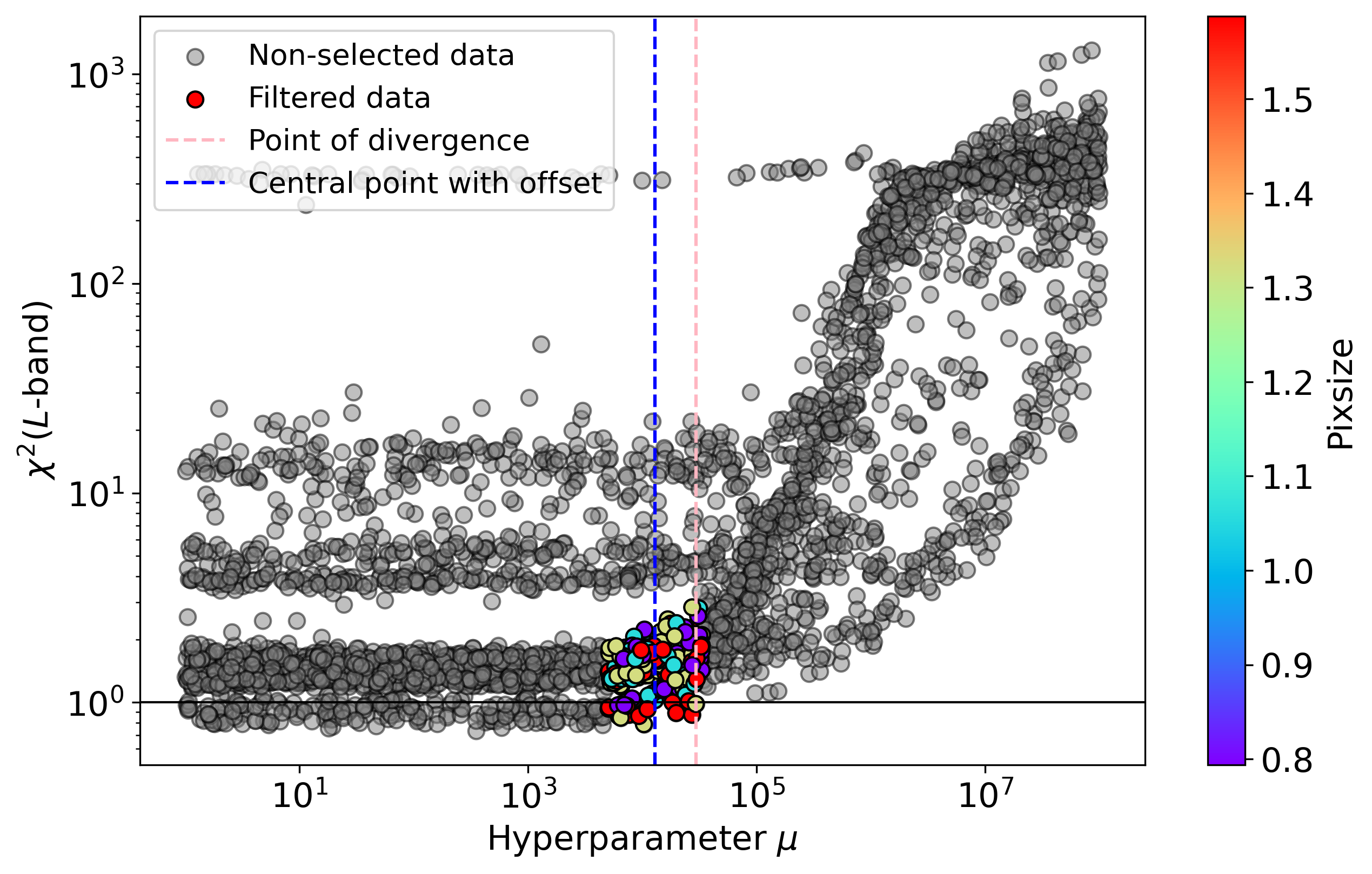}
    \caption{L-curves obtained for the reconstructed images of 3~Pup using VLTI/MATISSE $L$ band data. The colored points correspond to the solutions selected by \texttt{MYTHRA}.}
    \label{fig:l-curve-lPup_L}
\end{figure}

\begin{figure}[htbp!]
    \centering
    \includegraphics[width=0.75\linewidth]{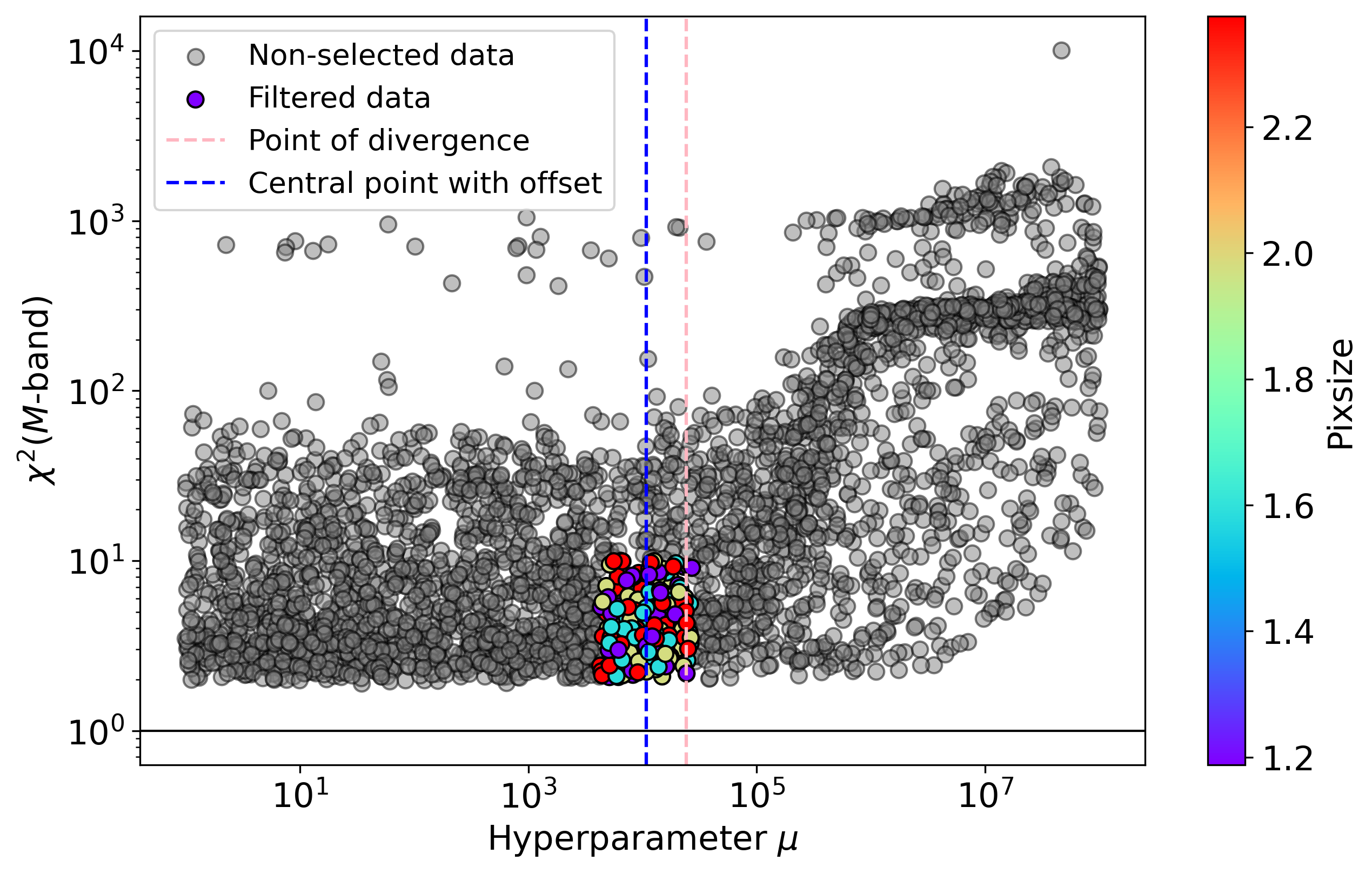}
    \includegraphics[width=0.75\linewidth]{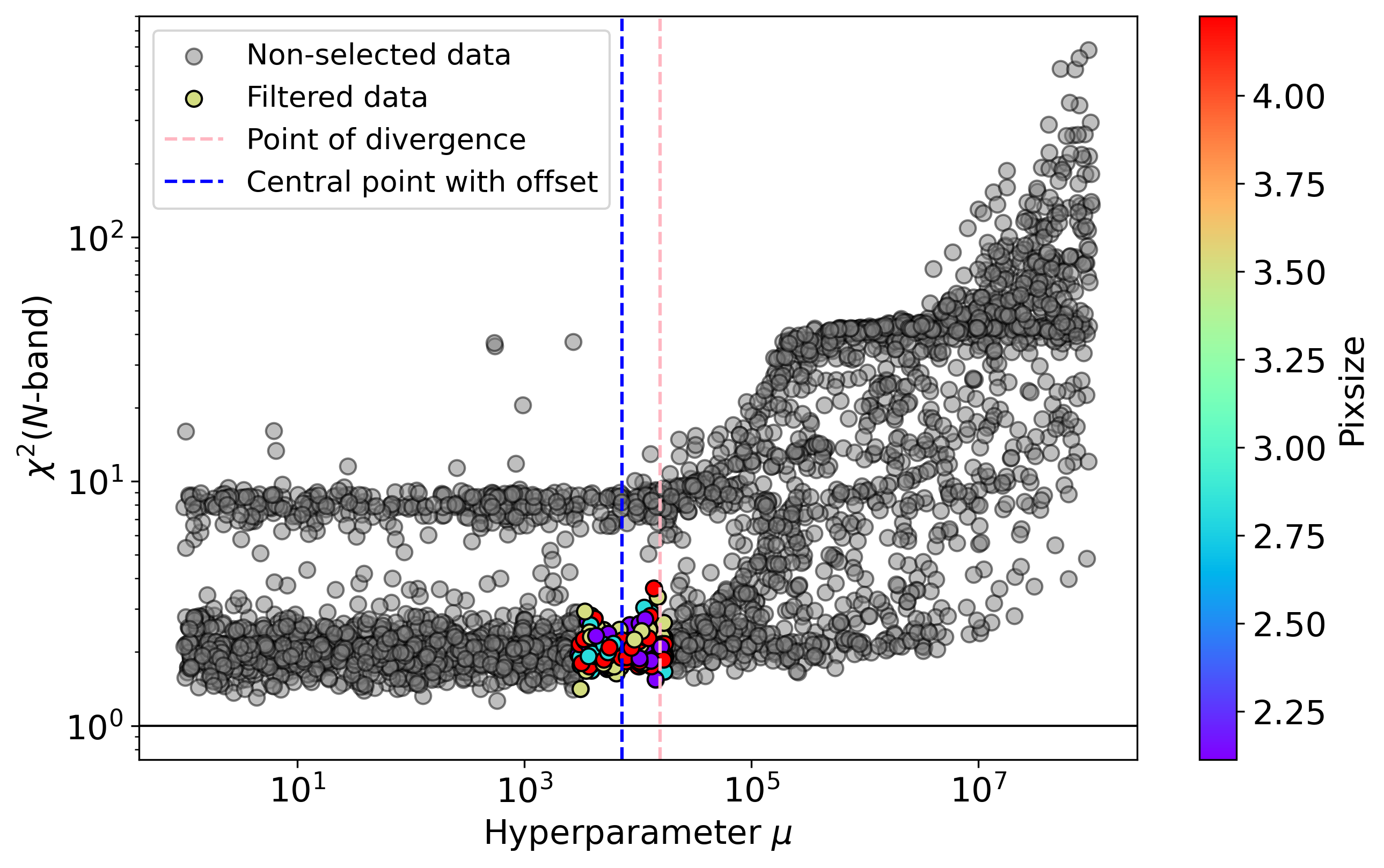}
    \caption{L-curves obtained for the reconstructed images of 3~Pup using VLTI/MATISSE data. The colored points correspond to the solutions selected by \texttt{MYTHRA}. Top: in the $M$ band. Bottom: in the $N$ band.}
    \label{fig:l-curve-lPup_MN}
\end{figure}

\begin{figure}[htbp!]
    \centering
    \includegraphics[width=0.75\linewidth]{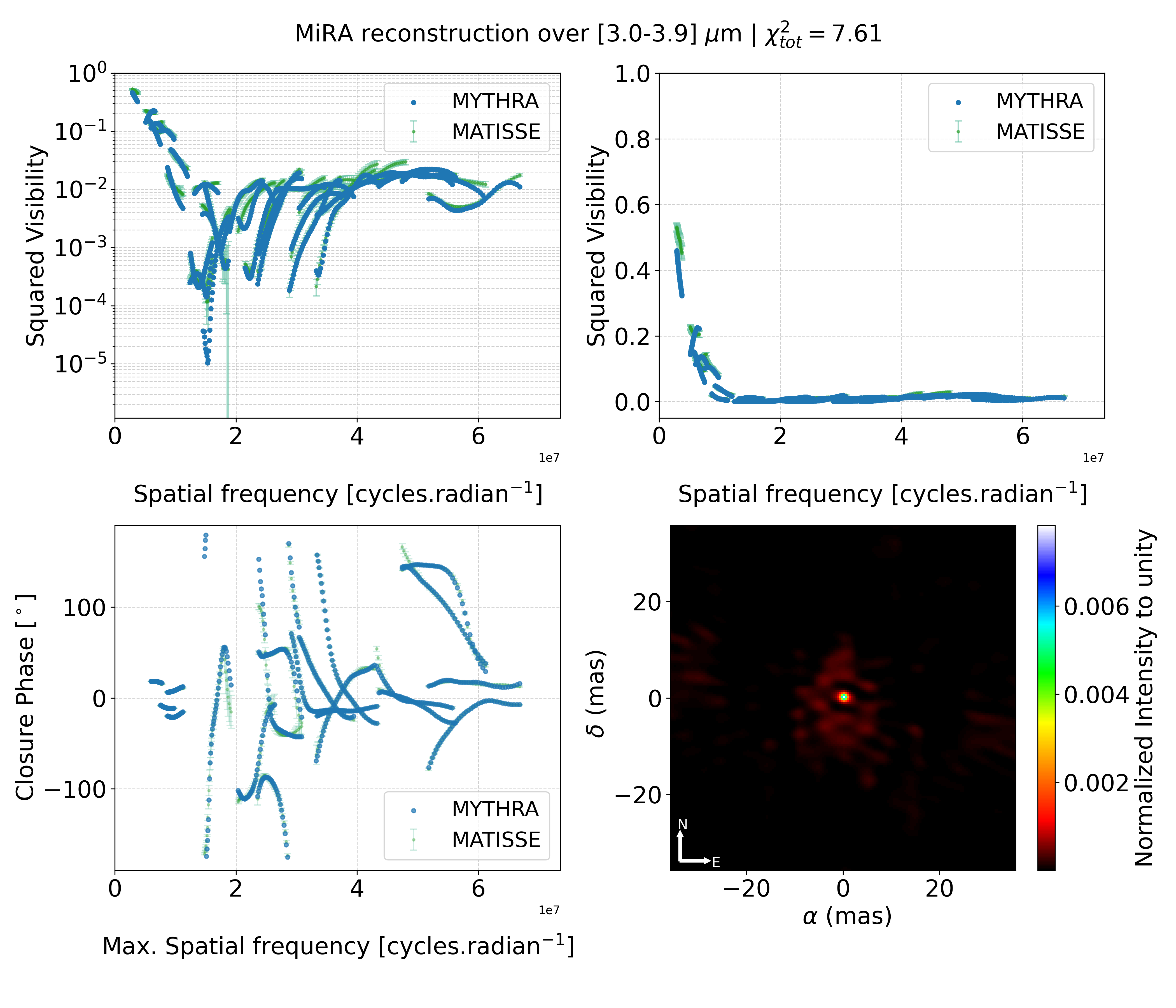}
    \includegraphics[width=0.75\linewidth]{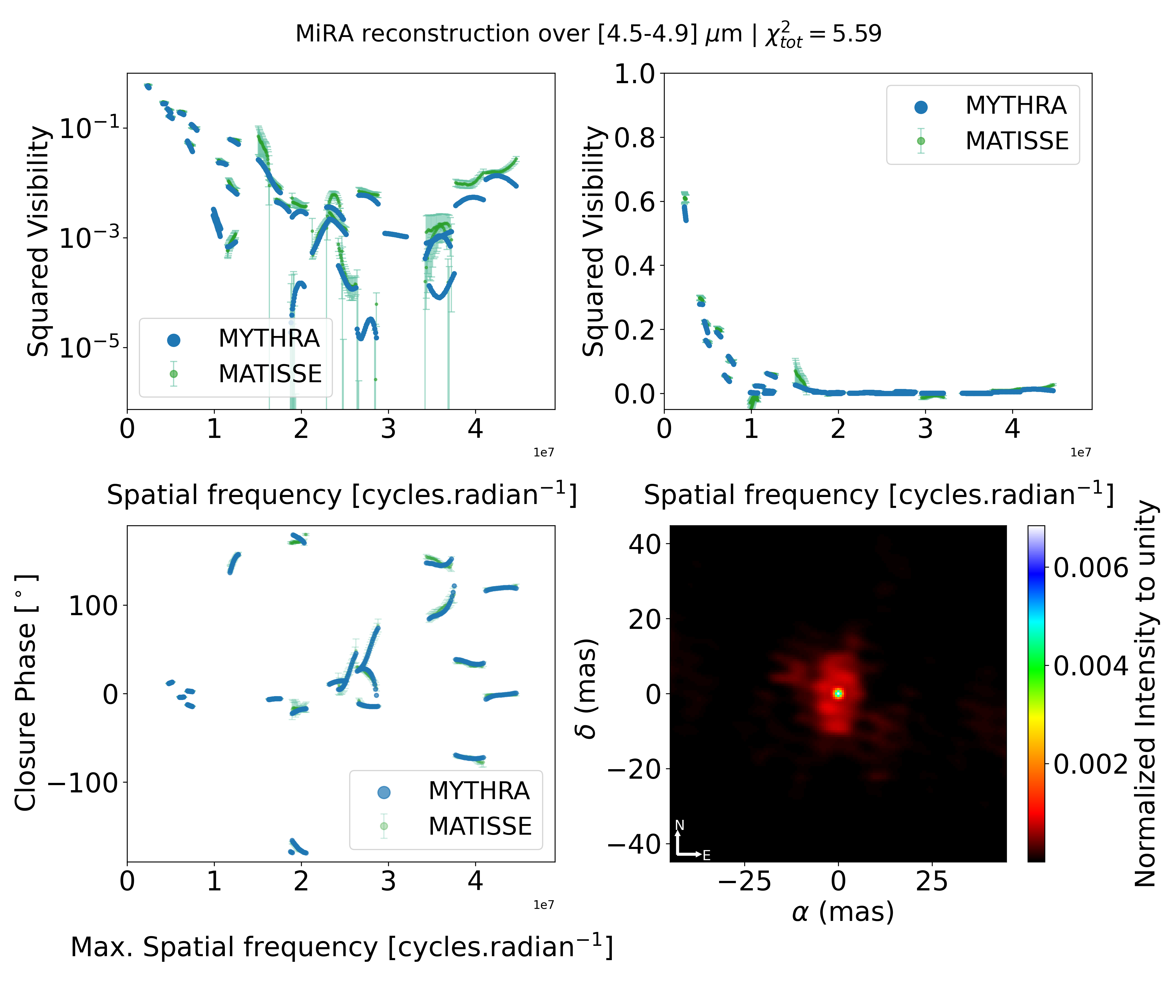}
    \caption{Images of 3~Pup reconstructed with \texttt{MYTHRA}, including the comparison with the observed squared visibilities (in logarithmic and linear scales) and the observed closure phases. Top: in the $L$ band. Bottom: in the $M$ band.}
    \label{fig:images_LM_lPup}
\end{figure}

\begin{figure}[htbp!]
    \centering
    \includegraphics[width=0.75\linewidth]{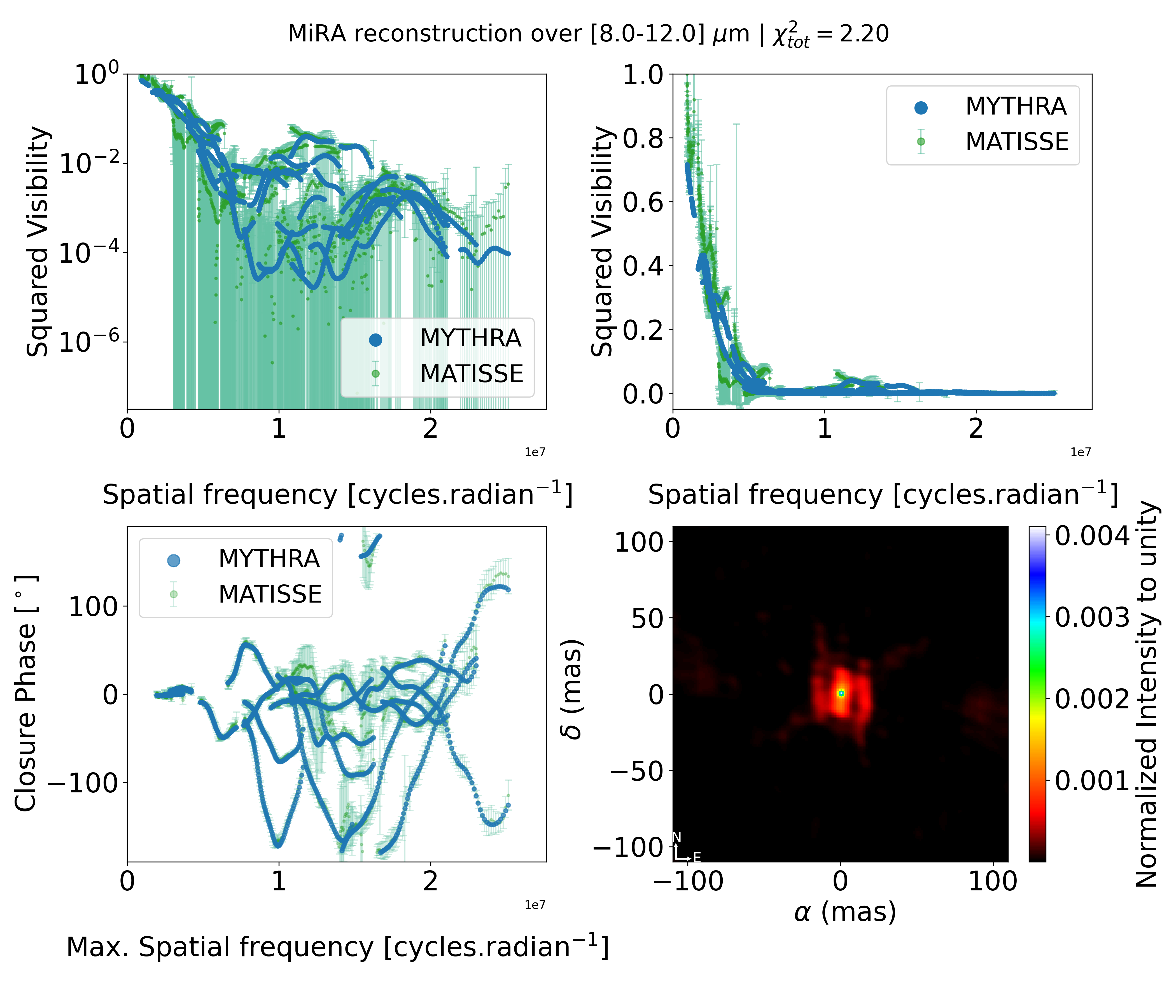}
    \caption{Images of 3~Pup reconstructed with \texttt{MYTHRA}, including the comparison with the observed squared visibilities (in logarithmic and linear scales) and the observed closure phases in the $N$ band.}
    \label{fig:images_N_lPup}
\end{figure}

\begin{figure}[htbp!]
    \centering
    \includegraphics[width=0.7\linewidth]{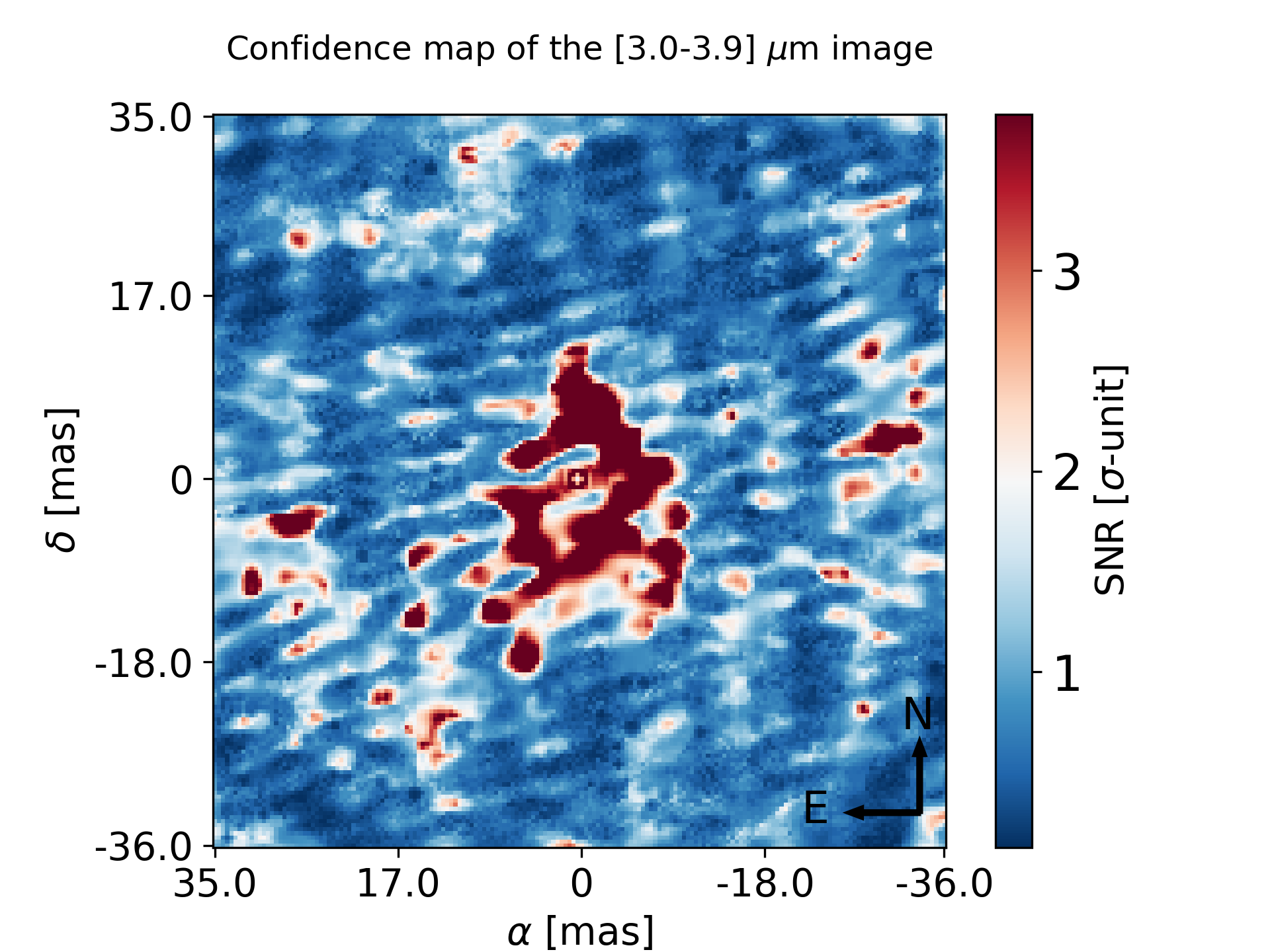}
    \caption{Error maps of the three averaged images reconstructed from 3~Pup VLTI/MATISSE $L$ band data using the hyperbolic regularization method.}
    \label{fig:confmaps_lPup_Lband}
\end{figure}

\begin{figure}[htbp!]
    \centering
    \includegraphics[width=0.7\linewidth]{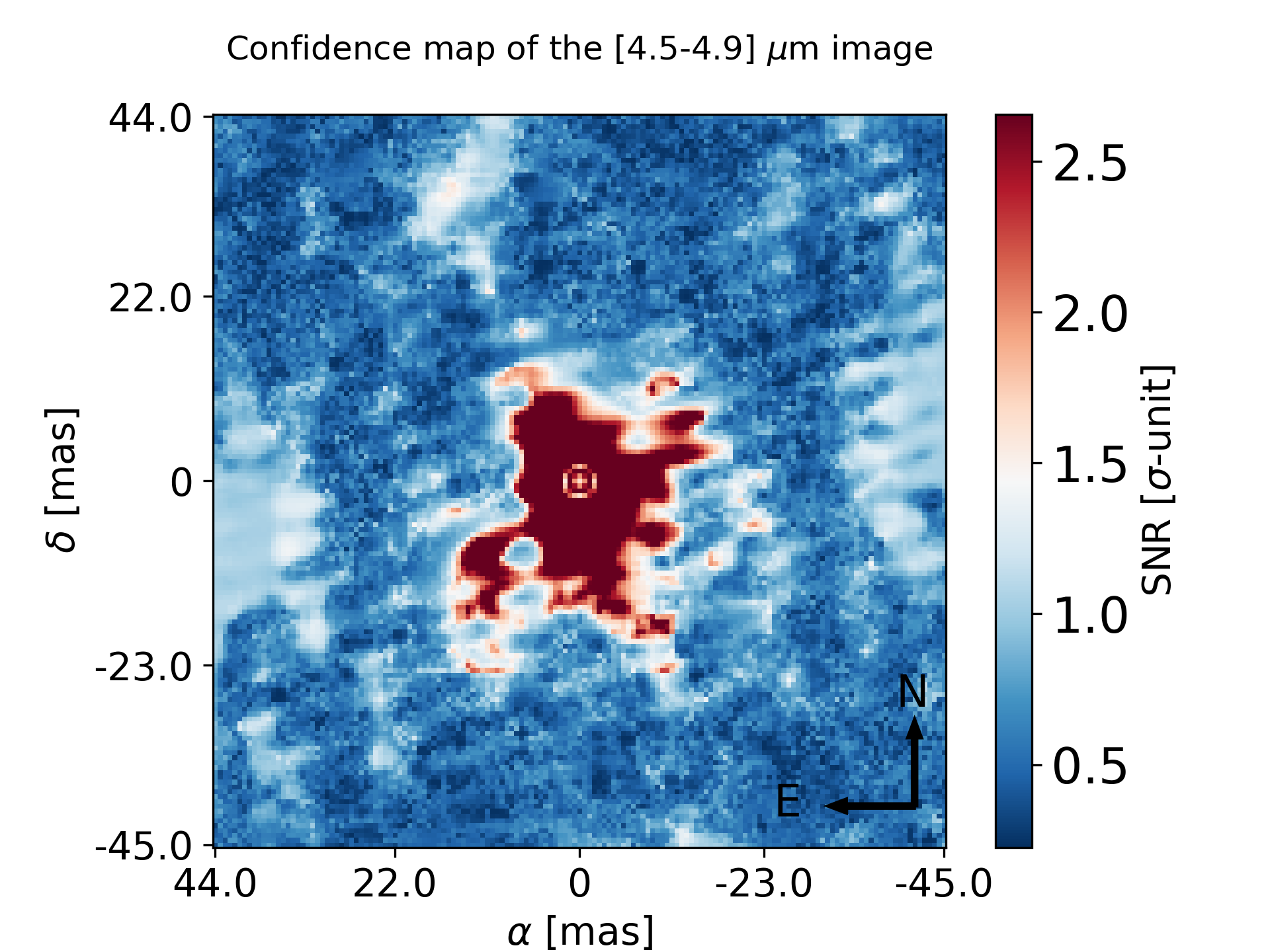}
    \includegraphics[width=0.7\linewidth]{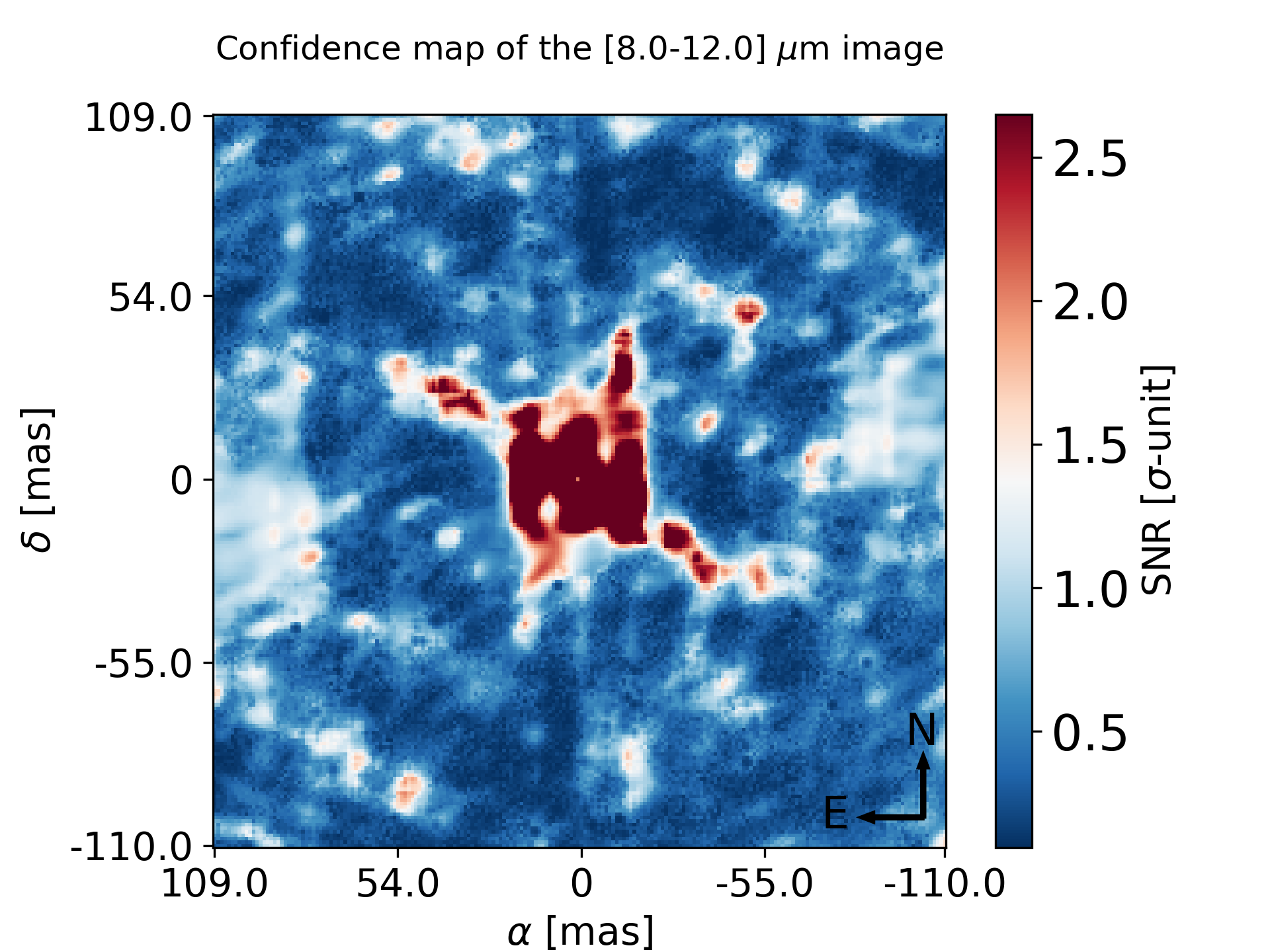}
    \caption{Error maps of the three averaged images reconstructed from 3~Pup VLTI/MATISSE data using the hyperbolic regularization method. Top: in the $M$ band. Bottom: in the $N$ band.}
    \label{fig:confmaps_lPup_MNbands}
\end{figure}

\begin{figure}[htbp!]
    \centering
    \includegraphics[width=0.7\linewidth]{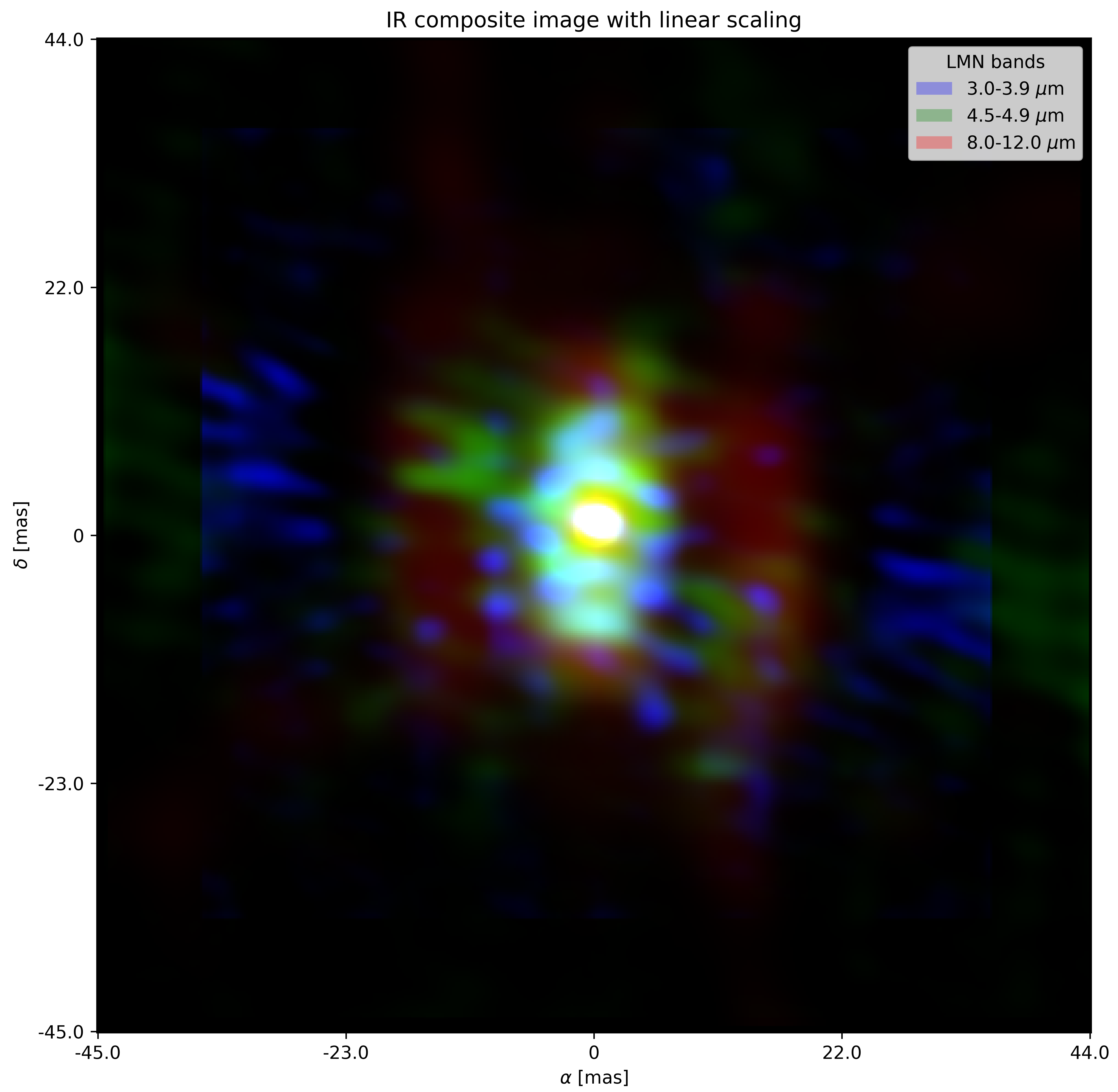}
    \caption{The RGB composite image of 3~Pup combines the three images reconstructed with \texttt{MYTHRA} in the $L$, $M$, and $N$ bands from VLTI/MATISSE data.}
    \label{fig:composite_lPup}
\end{figure}

\end{document}